\documentclass{iopart}
\usepackage{iopams}  
\usepackage{graphicx}
\usepackage{epsfig}

\newcommand{\be}{\begin{equation}}
\newcommand{\ee}{\end{equation}}
\newcommand{\bea}{\begin{eqnarray}}
\newcommand{\eea}{\end{eqnarray}}

\newcommand{\bm}{\mathbf}

\begin{document}

\title{What is the temperature of a granular medium?}

\author{A. Baldassarri\dag, A. Barrat\ddag, G. D'Anna\S,\\
V. Loreto\dag, P. Mayor\S and A. Puglisi\ddag
}
\address{\dag Physics Department, ``La Sapienza'' University and
INFM-SMC, P.le A. Moro 5, 00185 Rome, Italy}

\address{\ddag Laboratoire de Physique Th\'eorique (UMR du CNRS 8627),
B\^atiment 210, Universit{\'e} de Paris-Sud 91405 Orsay, France}

\address{\S Institut de Physique de la Mati\`ere Complexe, Facult\'e 
des Sciences de Base, Ecole Polytechnique F\'ed\'erale de Lausanne, 
CH-1015 Lausanne, Switzerland}

\begin{abstract}
In this paper we discuss whether thermodynamical concepts and in
particular the notion of temperature could be relevant for the
dynamics of granular systems. We briefly review how a temperature-like
quantity can be defined and measured in granular media in very
different regimes, namely the glassy-like, the liquid-like and the
granular gas. The common denominator will be given by the
Fluctuation-Dissipation Theorem, whose validity is explored by means
of both numerical and experimental techniques. It turns out that,
although a definition of a temperature is possible in all cases, its
interpretation is far from being obvious. We discuss the possible
perspectives both from the theoretical and, more importantly, from the
experimental point of view.
\end{abstract}

\maketitle

\section{Introduction}

Granular matter~\cite{Nagel,Nagel2,volume_ucgmg} constitutes one of
the most famous example of non-equilibrium, athermal systems in which
the usual temperature does not play any role.  However, the fact that
they involve many particles brings naturally a strong motivation to
treat them with thermodynamic or thermodynamic-like methods. It is
therefore of primary importance to determine whether such attempts are
feasible, and if usual tools of thermodynamics can be generalized. In
particular, can concepts like entropy or temperature have any meaning
or use ?

A thermodynamic approach is in general justified when one is able to
identify a distribution that is left invariant by the dynamics ({\em
e.g.} the microcanonical ensemble), and to assume that this
distribution will be reached by the system, under suitable conditions
of `ergodicity'.  Unfortunately, because energy is lost through
collisions or internal friction, and gained by a non-thermal source
(vibrations, tapping, shearing,...), the dynamical equations do not
leave the microcanonical or any other known ensemble invariant.  This
raises several questions, among which the most important is whether it
is possible in principle to construct a coherent thermodynamics for
these ``non-thermal'' systems.

The concept of temperature has been in fact widely used in the context
of the different states of granular media. The most straightforward
definition comes from the case of dilute, strongly vibrated granular
systems, which reach a non-equilibrium stationary state: by analogy
with molecular gases, a ``granular temperature'' $T_g$ can be defined
in terms of the average local kinetic energy per particle, and treated
as one of the hydrodynamic fields. This approach can a priori be
extended to denser, liquid-like, strongly vibrated systems.

In dense, slowly evolving dense granular media, the situation is more
complex since reaching a stationary state is experimentally or
numerically very difficult. These systems actually exhibit
aging~\cite{nicodemi,response} and memory~\cite{joss,memory}.
Analogies with other aging systems have also led to the definition of
dynamic temperatures as quantifying the violation of the equilibrium
Fluctuation-Dissipation Theorem (FDT)~\cite{FDT_general,FDT_general_2}.

In this paper, we will briefly review the different cases of dense,
liquid-like and gas-like granular media, focusing on the notion of
temperature as defined in the framework of the Fluctuation-Dissipation
Theorem.

The paper is organized along the following lines.  In section
\ref{section:fdt} we review the Fluctuation-Dissipation Theorem and
its generalizations in order to provide the tools that will be used to
test its validity in the different regimes.  In Section
\ref{section:dense} we discuss the case of dense granular media. We
first present a simple pedagogical $1$-d model for a granular medium
where the notions of entropy and temperature emerge naturally and that
could be used as a reference, although unrealistic, example. We then
review the basics of Edwards' approach as well as some results
obtained recently to test its validity. Next, in section
\ref{section:liquid} we describe experiments devoted to the test of
the Fluctuation-Dissipation Theorem in a vibro-fluidized granular
medium, probed by means of a torsion pendulum.  Section
\ref{section:gas} is devoted to the discussion of the case of granular
gases. Also in this case we shall present some procedures to test the
validity of the Fluctuation-Dissipation Theorem and to define a notion
of temperature.  Finally Section \ref{section:conclusion} is devoted
to the conclusions as well as to drawing some perspectives.

\section{Fluctuation-Dissipation Theorem and generalizations}
\label{section:fdt}

Let us consider an equilibrium system in contact with a thermostat at
temperature $T$.  For the purposes of this paper, the
Fluctuation-Dissipation Theorem~\cite{FDT_general} can be seen as
relating the random diffusion and the mobility of a tracer particle in
a gas or liquid: one possible version is the Einstein relation
$\left\langle [ r(t)-r(t') ]^2 \right\rangle = 2 d T \frac{\delta
\left\langle r(t) -r(t') \right\rangle }{\delta f}$, where $r$ is the
position of the particle and $f$ is a constant perturbing field, and
the brackets denote average over realizations.

Since the FDT is a feature of equilibrium systems, nothing guarantees
its validity in out of equilibrium systems. The situation is even 
worse in athermal systems, since it is not clear which quantity 
should play the role of the temperature.

Recent developments in glass theory, especially those related to their
out of equilibrium dynamics, have shown that in slowly evolving, aging
systems, the FDT is in fact modified in a very interesting fashion
(for a recent review see~\cite{reviewCrisantiRitort}).  In a class of
mean-field models, which contains, although in a rather schematic way,
the essentials of glassy phenomena~\cite{KTW,KTW2}, and whose aging
dynamics was solved analytically~\cite{CuKu}, a feature that emerged
was the existence of a temperature $T_{dyn}$ for {\em all} the slow
modes (corresponding to structural rearrangements)
\cite{CuKuPe,review}. This dynamical temperature $T_{dyn}$ in fact
exactly replaces the temperature of the heat bath in the Einstein
relation; it is different from the external temperature, but it can be
shown to have all other properties defining a true
temperature~\cite{CuKuPe}. In particular, the measure of $T_{dyn}$ can
be done using any version of the FDT relating the correlation and
response of an observable \cite{berthier:2002}.  Subsequently, the
violation of FDT has become a widely studied tool in the context of
glassy dynamics~\cite{CugliaHouches,reviewCrisantiRitort}.

The analogies between glassy thermal systems and non-thermal systems
close to jamming~\cite{Liu} have stimulated the investigations about
the existence of dynamical or effective temperatures in non-thermal
systems.  In particular, the appearance of a dynamical temperature in
models for dense, compacting granular media has been shown to arise
from an Einstein-like relation
\cite{bkls,bkls2,vittoria,vittoria2,makse}. We will focus on this aspect
in section~\ref{section:dense}.

Other works on dense granular matter have focused on another version
of the Fluctuation-Dissipation Theorem which relates the energy
fluctuations to the heat capacity~\cite{brey,lefevredean2001,fierro}.
Moreover,
another athermal system has been investigated by Ono et
al.~\cite{Ono:2002}. With reference to a model of sheared foam,
various possible definitions of effective temperature have been shown
to coincide, in particular in the context of an Einstein-like relation
and of energy fluctuations.

On another side of the wide range of non-equilibrium systems, granular
gases are very far from glassy systems. They are maintained in a
dilute non-equilibrium steady-state in which the dissipation due to
inelastic collisions between particles is compensated by a strong
energy injection. However, the existence of an Einstein relation
between diffusion and mobility of a tracer particle has also been
investigated recently in this context
\cite{Dufty:2001,Puglisi:2002,Barrat:2004,Garzo:2004}, as will be
developed in section~\ref{section:gas}.

It should be remarked that up to now most of the work in this area has
been theoretical and numerical and few experiments have been carried
out in order to check the validity of the Fluctuation-Dissipation
Theorem in real granular media. One example in this direction is
discussed in section~\ref{section:liquid}, concerning a
vibro-fluidized medium in a liquid-like regime~\cite{dannanature}.

\section{Dense granular media}
\label{section:dense}

\subsection{A pedagogical $1-d$ model}

In this section we consider a simple model which describes the
evolution of a system of particles which hop on a lattice of
$k=0,...,N$ stacked planes, as introduced in~\cite{prltetris} and
discussed in~\cite{prlema}.  In particular the system represents an
ensemble of particles which can move up or down in a system of $N$
layers in such a way that their total number is conserved.  We ignore
the correlations among particles rearrangements and problems related
to the mechanical stability of the system.  The master equation for
the density $\rho_k$ on a generic plane $k$, except for the $k=0$
plane, is given by:
\begin{eqnarray}
\partial_t \rho_k &=& (1-\rho_k)D(\rho_k) [ \rho_{k-1} \cdot p_u
+\rho_{k+1} \cdot p_d) ]+\\ && -\rho_k [(1-\rho_{k-1}) D(\rho_{k-1})
p_d +(1-\rho_{k+1}) D(\rho_{k+1})p_u] \nonumber
\label{n-plan}
\end{eqnarray}
where $p_d$ and $p_u$ (with $p_u+p_d=1$) represent the probabilities
for the particles to move downwards or upwards, respectively, among
the different planes. $D(\rho_{k})$ represents a sort of mobility for
the particles given by the probability that the particle could find
enough space to move.  Apart from other effects it mainly takes into
account the geometrical effects of frustration, i.e.  the fact that
the packing prevents the free movement of the particles.  With $p_u$
and $p_d$ we can define the quantity $x=p_u/p_d$ which quantifies the
importance of gravity in the system. We can associate to $x$ a
temperature for the system given by $T \sim 1 / log(1/x)$.

One interesting question to address is whether there exists a
variational principle driving the relaxation phenomena in this system
and in general in granular media. In other words one could ask whether
some free-energy-like functional is minimized (Lyapunov functional)
~\cite{Lyapunov} by the dynamical evolution. In this model, and more
generally for evolution equations of the form
\begin{equation}
\partial_t \rho_k = g(\rho_k) [ f(\rho_{k-1}) p_u + f(\rho_{k+1}) p_d)
] -f(\rho_{k}) [g(\rho_{k-1}) p_d + g(\rho_{k+1}) p_u]
\label{dymogen}
\end{equation} 
where $f$ and $g$ are generic functions for which we only require $f
\ge 0$, $g \ge 0$, $ d f/ d\rho \ge 0$, $ d g /d \rho \le 0$,
it is
possible to prove that such a functional, which decreases
monotonically along the trajectories of the motion, indeed exists and
is given by:
\begin{equation}
F=\sum_{k=0}^{\infty} [ \gamma(x) k \rho_k - S(\rho_k) ] \ \ ,
\label{funcmonogen}
\end{equation}
with $S(\rho_k) = \int_{\rho_k} \log {g(\rho)/ f(\rho)} d\rho$ and
$\gamma(x) = \log(1/x)$.

A deeper insight in the above mentioned phenomenology is obtained by
considering the continuum limit for the model described by
(\ref{dymogen}).  More precisely we consider a diffusive limit that
consists in scaling the space variable as $1\over \epsilon$, the time
variable as ${1\over \epsilon^{2}}$ and the drift term $p_d-p_u$ее as
$\epsilon$.  Therefore $x=\epsilon k$, $\tau=\epsilon^{2}k/2$,
$p_d-p_u=\epsilon{\beta}/2$ее and we consider the evolution of
$u(x)\equiv\rho(k)$.  We get the continuum limit by taking the Taylor
expansion of the right hand side of (\ref{dymogen}) around
$x=k\epsilon$. For example $\rho(k+1)\equiv
u(x+\epsilon)=u(x)+\epsilon\partial_{x}uе+ {1\over
2}\epsilon^{2}\partial_{xx} u+O(\epsilon^{3})$.  We get, formally, \be
\partial_{\tau}u(x)=\beta\partial_{x}е(f g) + \left(g \partial_{xx}f -
f \partial_{xx}еg\right) + O(\epsilon), \ee which, in the limit
$\epsilon\rightarrow 0$, gives \be
\partial_{\tau}u(x)=\beta\partial_{x}е(f g) +(g \partial_{xx}еf -f
\partial_{xx}еg) .
\label{continuum1}
\ee 

\noindent This non-linear diffusion equation may be conveniently written in the
following form \be \partial_{\tau}u=\partial_{x}\left(D(u)ее
\partial_{x}{\partial F\over\partial_{u}}\right),
\label{continuum2}
\ee 

\noindent where $D=f g$, ${\partial F\over\partial_{u}}$ denotes the
functional derivative of $F$ with respect to $u$, $F=\int_0^{\infty}
(\beta\phantom{,} u\phantom{,}x - S(u)) dx,$е and
$S'=\log\left({g\over f}\right)$.  Notice that the functional $F$
decreases with the dynamics induced by (\ref{continuum1}). One
has, in fact, \be \partial_{\tau}F=\int dx {\partial F\over\partial u
}\partial_{\tau}u=\int dx {\partial F\over\partial u
}\partial_{x}\left(D(u)ее \partial_{x}{\partial F\over\partial u
}\right)
\label{cahn}
\ee 
\noindent that, after an integration by parts, gives \be -D(u)
\left({\partial F\over\partial_{u}}\right)^{2}\leq 0.  \ee Therefore
there exists a ``free energy''-like functional, $F$, for
(\ref{continuum2}) which has exactly the same form of the
functional defined for the discrete model (see
(\ref{funcmonogen})).  We can notice that, while the functional
form of $S$ and the value of $\beta$ determine in a unique way the
asymptotic state, they are not sufficient to determine the dynamical
behavior of the system.  In particular in order to know it one should
know the functional form of $D(\rho)$.

What the analysis of this simple model suggests is the possibility of
introducing, for non-thermal systems such as granular media, equilibrium
concepts like free-energy, entropy and temperature.  More precisely it
is possible (in the case studied here) to predict the asymptotic state
by means of the minimization of a suitable functional which can be
constructed by entropic arguments.  It is worth stressing how granular
systems often exhibit memory and so the existence of a unique Lyapunov
functional is not guaranteed in general.  One could for example expect
that several Lyapunov functionals are associated to different
stationary states reached with different dynamical paths.

\subsection{Edwards' approach}

A very ambitious approach was put forward some years ago by
S.F. Edwards and collaborators \cite{Sam,edw_glasses,anita,Repo}, by
proposing for dense granular an equivalent of the microcanonical
ensemble. The idea is to suggest that one could reproduce the
observables attained dynamically by first measuring the density of the
system, and then calculating the value of the observables in an
ensemble consisting of all the {\em `blocked' configurations} at the
measured density. The blocked configurations are defined as as those
in which every grain is unable to move. 

This `Edwards ensemble' leads naturally to the definition of an
entropy $S_{Edw}$, given by the logarithm of the number of blocked
configurations of given volume, energy, etc., and its corresponding
density $s_{Edw}\equiv S_{Edw}/N$.  Associated with this entropy are
state variables such as `compactivity'
$X_{Edw}^{-1}=\frac{\partial}{\partial V}S_{Edw}(V)$ and `temperature'
$T_{Edw}^{-1}=\frac{\partial}{\partial E}S_{Edw}(E)$.

That configurations with low mobility should be relevant in a jammed
situation is rather evident, the strong hypothesis here is that the
configurations reached dynamically are {\em the typical ones} of given
energy and density. Had we restricted averages to blocked
configurations having {\em all} macroscopic observables coinciding
with the dynamical ones, the construction would exactly, and
trivially, reproduce the dynamic results. The fact that conditioning
averages to the observed energy and density suffices to give other
dynamical observables (even if maybe only as an approximation) is
highly non-trivial.

It turns out that the advances in glass theory mentioned in
section~\ref{section:fdt} have in fact come to clarify and support
such a hypothesis.  Indeed, the dynamical temperature emerging from
the Einstein-like relation between diffusion and mobility, despite its
very different origin, matches exactly Edwards' ideas.  One can indeed
identify in mean-field models all the energy minima (the blocked
configurations in a gradient descent dynamics), and calculate
$1/T_{Edw}$ as the derivative of the logarithm of their number with
respect to the energy. An explicit computation shows that $T_{Edw}$
coincides with $T_{dyn}$ obtained from the out of equilibrium dynamics
of the models aging in contact with an almost zero temperature
bath~\cite{remi,jamming,Theo,Frvi,Felix,biroli}.  Moreover, given the
energy $E(t)$ at long times, the value of any other macroscopic
observable is also given by the flat average over all blocked
configurations of energy $E(t)$.  Within the same approximation, one
can also treat systems that, like granular matter, present a
non-linear friction and different kinds of energy input, and the
conclusions remain the same~\cite{jorge-trieste} despite the fact that
there is no thermal bath temperature.
 
Edwards' scenario then happens to be correct within mean-field schemes
and for very weak vibration or forcing. The problem that remains is to
what extent it carries through to more realistic models.  In the next
subsections, we present the general methodology that has been used to
explore this issue for some representative examples of models of
granular compaction.

\subsection{General strategy to check Edwards' assumption:}

One possibility of making an assumption {\em \`a la} Edwards would be to
consider a fast quench, and then propose that the configuration
reached has the macroscopic properties of the typical blocked
configurations. This would imply that the system stops at a density
for which the number of blocked configurations is maximal.  However,
it turns out that generically the vast majority of the blocked
configurations are much less compact than the one reached dynamically,
even after abrupt quenches.

One has therefore to give up trying to predict the dynamical energy or
density by methods other than the dynamics itself. The strategy here
is to quench the system to a situation of very weak tapping, shearing
or thermal agitation. In this way, the system keeps compactifying,
albeit at a very slow rate. Edwards' measure is then constructed as a
flat measure over blocked configurations {\em conditioned to have the
energy and/or density of the dynamical situation}.

In order to check Edwards' hypothesis, we will show how Edwards'
measure can be constructed in some representative (non mean-field)
systems, together with the corresponding entropy and expectation
values of observables. Moreover, we also construct what can be called
Gibbs' measure by removing the constraint of sampling blocked
configurations. Both measures are then compared with the observables
obtained with an irreversible compaction dynamics.

\subsubsection{Models}

To illustrate the strategy to check Edwards' hypothesis we shall
mainly refer to one specific finite-dimensional lattice model which
has been shown to reproduce the complex phenomenology of the granular
compaction. We focus on the so-called Kob-Andersen (KA)
model~\cite{KoAn}, which considers a three-dimensional lattice gas
with at most one particle per site and otherwise purely dynamical
constraints: a particle can move to a neighboring empty site, only if
it has strictly less than $m$ neighbors in the initial and in the
final position.  This model was first devised in the context of
Mode-Coupling theories to reproduce the cage effect existing in
supercooled liquids, which produces at high density a very strong
increase of the relaxation time\footnote{For recent and interesting
results about this model the reader is referred
to~\cite{BiroliToninelli}}.  Though very schematic, it has then been
shown to reproduce rather well several aspects of
glasses~\cite{KuPeSe,Se} and of granular compaction~\cite{SeAr}.  The
simplicity of its definition and the fact that it is non mean-field
makes it a very good candidate to test Edwards' ideas: in fact, the
triviality of its Gibbs measure allows to compare the numerical data
obtained for the dynamics and for Edwards' measure with the analytic
results for equilibrium.

Completely similar results have been obtained in the framework of another
class of non mean-field models, the so-called Tetris Model
\cite{prltetris,response}. Here the constraints are not purely
dynamical, but related to the steric properties of the grains which
undergo a geometrical (and hence dynamical) frustration. We refer the
reader to~\cite{bkls2,vittoria,vittoria2} for details.

\subsubsection{Equilibrium measure}

Let us first consider the case of the Kob-Andersen model: the dynamic
character of the rule guarantees that the equilibrium distribution is
trivially simple since all the configurations of a given density are
equally probable: the Hamiltonian is just $0$ since no static
interaction exists. Density and chemical potential are related by
$\rho = 1/(1+\exp(- \beta \mu))$, and the exact equilibrium entropy
density per particle reads
\begin{equation}
  s_{equil}(\rho) = -\rho \ln \rho - (1-\rho)\ln(1-\rho) \,\,
  \rightarrow \frac{ds_{equil}}{d\rho} = - \beta \mu.
  \label{eq:s_equil}
\end{equation} 

\noindent In this model, the temperature $1/\beta$, which
appears only as a factor of the chemical potential, is irrelevant, so that we
can set it to one. Besides, the equilibrium structure factor is easily seen to
be a constant, $g_{equil}(r) = \rho^2$: no correlations appear since the
configurations are generated by putting particles at random on the lattice. It
will therefore be easy, as already mentioned, to compare small deviations from
$g_{equil}(r)$, a notoriously difficult task to do in glassy systems.

\subsubsection{Edwards' measure}

Since Edwards' measure considers blocked configurations in which no
particle is allowed to move, the crucial step to sample this measure
is in fact to introduce an {\em auxiliary model}~\cite{bkls}: the
auxiliary energy $E_{aux}$ is defined as {\em the number of mobile
particles}, where a particle is defined as mobile if it can be moved
according to the dynamical rules of the original model. Edwards' measure
is thus a flat sampling of the ground states ($E_{aux}=0$) of this
auxiliary model, which is obtained by an annealing procedure, at fixed
density, of the auxiliary temperature $T_{aux}$ (and we write
$\beta_{aux}=1/T_{aux}$). Note that the Monte-Carlo dynamics of the
auxiliary model does not need to respect the constraints of the real
model, so that efficient samplings with e.g. non-local moves can be
obtained.

\begin{figure}[htb]
\centerline {
\includegraphics[clip=true,width=5cm,keepaspectratio]{entro_edw.eps}
\includegraphics[clip=true,width=6.0cm,keepaspectratio]{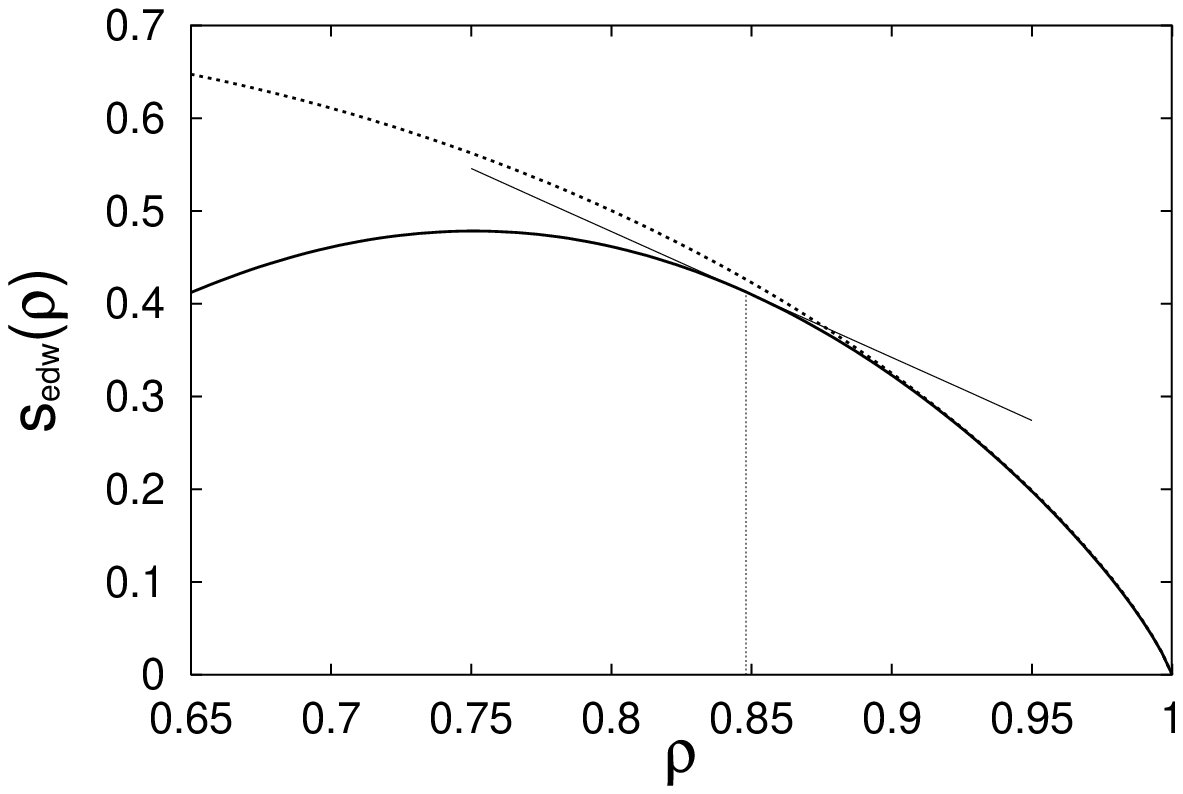}
}
\caption{Edwards' entropy density, $s_{Edw}(\rho)$ and equilibrium
entropy density, $s_{equil}(\rho)$ for the Tetris (left)
(see~\cite{bkls2} for details) and the Kob-Andersen (right) models.}
\label{entropia_edw}
\end{figure}

From the measure of the auxiliary energy during the annealing, at
given density $\rho$, $E_{aux}(\beta_{aux},\rho)$, one can compute the
Edwards' entropy density defined by:
\begin{equation}
s_{Edw}(\rho) \equiv s_{aux}(\beta_{aux}=\infty,\rho)= s_{equil}(\rho)
- \int_0^\infty e_{aux}(\beta_{aux},\rho) d\beta_{aux} \,
\label{form_entro_edw}
\end{equation}
where $e_{aux}(\beta_{aux},\rho)$ is the auxiliary Edwards' energy
density and $s_{equil}(\rho)= s_{aux}(\beta_{aux}=0,\rho)$.
Fig.~\ref{entropia_edw} reports the results for $s_{Edw}(\rho)$ as
obtained from (\ref{form_entro_edw}) compared with $s_{equil}(\rho)$,
for the Tetris (left panel) and the Kob-Andersen models (right panel),
respectively.

The slope of the tangent to $s_{Edw}(\rho)$ for a generic $\rho$
allows to extract $T_{Edw}(\rho)$.  The natural definition for
Edwards' temperature is
\begin{eqnarray}
T_{Edw}^{-1} &=& -\frac{1}{\mu} \frac{ds_{Edw}(\rho)}{d\rho} \, ;
\end{eqnarray}
which yields:
\begin{eqnarray}
  T_{Edw}(\rho) &=& {\frac{ds_{equil}(\rho)}{d\rho}} /
  {\frac{ds_{Edw}(\rho)}{d\rho}}.
\end{eqnarray}

Similarly, the Edwards measure structure function, $g_{Edw}(r)$, is
obtained as
\begin{eqnarray}
  g_{Edw}(r) & = & \lim_{\beta_{aux} \to \infty}
  g_{aux}(r,\beta_{aux}) \,.
\end{eqnarray}

\subsubsection{Irreversible Compaction Dynamics}

The irreversible compaction dynamics is obtained by trying to increase
the density of the system, starting from a low-density ``equilibrium''
situation. For the KA model, this can be done e.g. by increasing
slowly the chemical potential on a given layer of a three-dimensional
box: when the chemical potential becomes large enough, the dynamical
constraint does not allow anymore the system to reach the desired
density and slow compaction follows.

During the compaction, we record the density $\rho(t)$ and the density
of mobile particles $\rho_m(t)$.  It is particularly interesting to
notice that in the out-of-equilibrium configurations visited during
the irreversible dynamics, the fraction of mobile particles is
systematically smaller than the corresponding value in
equilibrium. This suggests the possibility of distinguishing between
equilibrium and out-of-equilibrium configurations by looking at the
spatial organization of the particles in both cases.  We have thus
measured, during the compaction dynamics, the particle-particle
correlation function at fixed density.

The existence of an Einstein relation during the compaction dynamics
is tested by the measure of the mobility of the particles
\be
\chi(t_w,t_w+t)=\frac{1}{dN}\sum_{a} \sum_{k=1}^{N} \frac{\delta
\left\langle (r_k^a(t_w+t) - r_k^a(t_w)) \right\rangle}{\delta f},
\ee
\noindent obtained by the application of random forces to the
particles between $t_w$ and $t_w+t$, and the mean square displacement

\be 
B(t_w,t_w+t)=\frac{1}{dN}\sum_{a} \sum_{k=1}^{N} \left\langle
(r_k^a(t_w+t)-r_k^a(t_w))^2 \right\rangle,
\ee
\noindent ($N$ is the number of particles; $a=1,\cdots,d$ runs over
the spatial dimensions: $d=2$ for Tetris, $d=3$ for KA; $r_k^a$ is
measured in units of the bond size $d$ of the square lattice).
Indeed, the quantities $\chi(t_w,t_w+t)$ and $B(t_w,t_w+t)$, at
equilibrium, are linearly related (and actually depend only on $t$
since time-translation invariance holds) by
\be
2 \chi(t) = \frac{X}{T_{eq}} B(t),
\label{FDT}
\ee
\noindent 
where $X$ is the so-called Fluctuation-Dissipation ratio (FDR) which
is unitary in equilibrium. Any deviations from this linear law signals
a violation of the Fluctuation-Dissipation Theorem (FDT). 
In particular, as mentioned in section~\ref{section:fdt},
in many aging systems, and in particular in the KA model~\cite{Se}, violations
from (\ref{FDT}) reduce to the occurrence of two regimes: a
quasi-equilibrium regime with $X=1$ (and time-translation invariance)
for ``short'' time separations ($t \ll t_w$), and the aging regime
with a constant $X \le 1$ for large time separations. This second
slope is typically referred to as a dynamical temperature $T_{dyn} \ge
T_{eq}$ such that $X = X_{dyn} = T_{eq}/T_{dyn}$~\cite{CuKuPe}.

\subsubsection{Comparing different measures}

At this stage it is possible to compare equilibrium and Edwards'
measures with the results of the out-of-equilibrium dynamics at large
times.

\begin{figure}[h]
\centerline
{
\includegraphics[clip=true,width=5.8cm,angle=0,keepaspectratio]{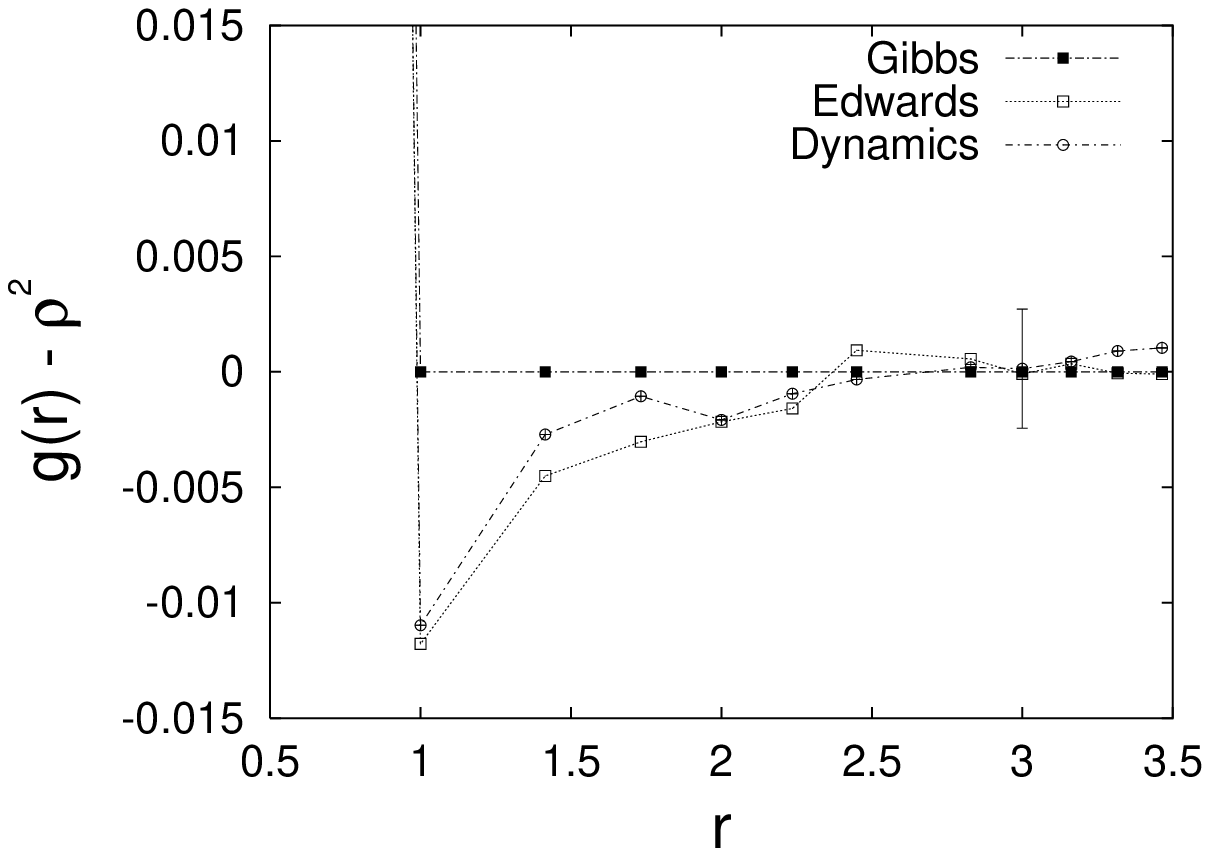}
\includegraphics[clip=true,width=5.0cm,angle=0,keepaspectratio]{correl_confr.eps}
}
\vspace*{8pt}
\caption{ {\bf Left: KA model} Dynamic structure function for the KA
model, obtained in a very slow compression ("Dynamics"), 
and with the equilibrium ("Gibbs") and Edwards' measure at the density
reached dynamically.
 {\bf Right: Tetris model} Comparison between the
correlation functions obtained with the equilibrium measure, the
Edwards measure ($\beta_{aux}=6$) and the irreversible dynamics. In
all cases the system is considered at a density of $\rho \simeq
0.58$.}
\label{correl_confr}
\end{figure}

In Fig.~\ref{correl_confr} we plot the deviations of the dynamical
particle-particle correlation functions from the uncorrelated value
$\rho^2$. In particular we compare $\langle
(g_{dyn}(r)-\rho^2)\rangle$ obtained during the irreversible
compaction with the corresponding functions obtained with the
equilibrium and Edwards' measures.  It is evident that the correlation
function, as measured during the irreversible compaction dynamics, is
significantly different from the one obtained with the equilibrium
measure. On the other hand the correlation functions obtained with
Edwards' measure are able to describe better what happens during the
irreversible dynamics. In particular what is observed is that the
correlation length seems to be smaller for configurations explored by
the irreversible dynamics than in the equilibrium configurations.
This aspect is captured by Edwards' measure which selects
configurations with a reduced particle mobility.  In practice one can
summarize the problem as follows: given a certain density, one can
arrange the particles in different ways. The different configurations
obtained in this way differ in the particle mobility and this feature
is reflected by the change in the particle-particle correlation
properties.

\begin{figure}[h]
\centerline
{
\includegraphics[clip=true,width=8cm,keepaspectratio]{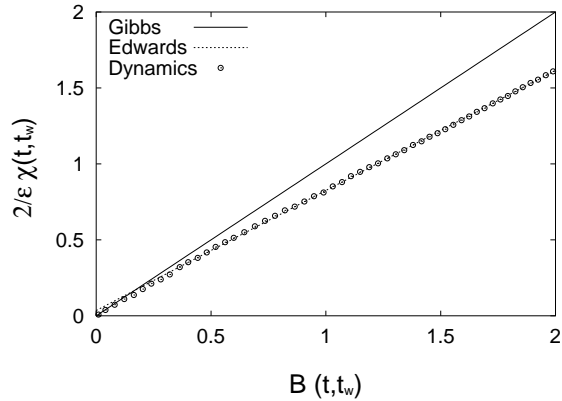}
}
\vspace{0.2cm}
\caption{Einstein relation in the Kob-Andersen model: plot of the
mobility $\chi(t,t_w)$ vs. the mean-square displacement $B(t,t_w)$
(data shown as circles).  The slope of the full straight line
corresponds to the equilibrium temperature ($T=1$), and the slope of
the dashed one to Edwards' prescription obtained from
figure~\ref{entropia_edw} at $\rho(t_w)=0.848$.}
\label{fig:x}
\end{figure}

Another comparison can be performed with regard to the dynamical
temperature $T_{dyn}$~\cite{Se}.  Figure~\ref{fig:x} shows a plot of
the mobility $\chi(t,t_w)$ {\em vs.} the mean square displacement
$B(t,t_w)$ testing in the compaction data the existence of a dynamical
temperature $T_{dyn}$~\cite{Se}. The agreement between $T_{dyn}$ and
the Edwards temperature $T_{edw}$, obtained from the blocked
configurations as in Fig.~\ref{entropia_edw}, for the density at which
the dynamical measurement were made, is clearly excellent.
Further evidences in this direction have been obtained for the Tetris
model~\cite{vittoria}.

To summarize, during the compaction, the system falls out of
equilibrium at high density, and is therefore no more described by the
equilibrium measure. It turns out that Edwards' measure, constructed
by a flat sampling of the blocked configurations at the dynamically
reached density, reproduces the physical quantities measured at large
times, and in particular predicts the correct value for the dynamical
temperature.

\subsection{Partial Conclusions} 

It turns out that Edwards' measure, constructed by a flat sampling of
the blocked configurations, is able to reproduce the physical
quantities measured at large times.
In particular, the connection of Edwards' temperature with the
dynamical FDT temperature seems to be generally valid (though there
are important counterexamples~\cite{bkls2,smedt}).

Apart from the first evidences reviewed here, which have lent credibility to
Edwards' construction, various works have allowed a better comprehension of
the validity of Edwards' approach and of its limitations. A comprehensive
review of these approaches is beyond the scope of the present paper. An
incomplete list would include results on $1$-d models of particles or
spins~\cite{brey,lefevredean2001,berg}, parking lot
models~\cite{tarjusviot2003}, $3$-d molecular dynamics
simulations~\cite{makse}, diluted spin-glasses and hard-spheres on
lattice~\cite{fierro}, spin-glasses with tapping~\cite{lefevredean2002}. 

It is important to note that, in a case for which the explicit analytical
computation of Edwards' measure and of the dynamical quantities is feasible,
it has been shown that Edwards' construction of a flat measure on the blocked
configurations is not exactly valid~\cite{smedt}. In~\cite{bkls2}, 
the study of the low-temperature dynamics of the Random Field Ising model
has also shown that the dynamically reached configurations have
typically zero magnetization, in contrast to the configurations dominating
Edwards' measure. While such studies show that Edwards' construction
should probably be considered only as a first approximation,
generalizations have been shown to yield
better and better results for the prediction of observables obtained
dynamically. These generalizations imply to use a flat average
on blocked configurations, restricted by constraining more than just
one variable to its value obtained dynamically
\cite{lefevre:2002,vittoria2}. The question arising concerns then the
number of observables to constrain in order to obtain reliable
predictions for the others.

While such theoretical aspects remain interestingly open, a crucial question
concerns the experimental validation of Edwards' ideas. In particular,
the study of diffusion and mobility of different tracer particles
within driven granular media would allow to confirm or disprove
the theoretically predicted violation of FDT and the existence
of dynamical temperatures
\footnote{For other alternative experimental tests for the
validity of Edwards' approach, inspired by the study of spin systems,
see~\cite{lefevredean2003}.}. 
Such experiments are actually
in progress and a first set of results concerning
the diffusion has been published in ~\cite{dauchot,clement2}.

\section{Liquid-like granular media} 
\label{section:liquid}

\subsection{Context}

In this section we analyze whether a notion of temperature can be
defined for a granular medium in a liquid-like regime, i.e. a regime
where the medium is brought by vibration to a quasi-fluidized state.
We shall mainly refer to some recent experimental results where a
vibrate granular medium was sensed by means of a torsion pendulum.

In the classical Brownian motion experiment, a ``tagged'' particle
immersed in a liquid can be used as a thermometer to determine the
temperature of the liquid itself. For this, one has to record the
motion of the tagged particle, and data analysis, for example
according to the Langevin formalism, gives the temperature.  One may
wonder whether a similar experiment, performed in a granular medium
under suitable external vibrations (so that it looks very much like a
liquid), could be used to determine a ``granular-liquid temperature".

\subsection{Theoretical background}


Let us briefly review for clarity the behaviour of
a torsion oscillator of moment of inertia $I$ and
elastic constant $G$ immersed into an equilibrium liquid.
Following the Langevin hypothesis~\cite{langevin}, we suppose that the
effect of this perturbing environment is split into two parts: a
viscous friction force, proportional to the oscillator angular
velocity, and a random, rapidly fluctuating force~$\xi(t)$, which is an
uncorrelated Gaussian white noise of zero mean and variance $q$.
The oscillator angular position $\theta$ satisfies the
 Langevin equation:
\begin{equation}
I\ddot{\theta}(t)+\alpha 
\dot{\theta}(t)+G \theta(t)=\xi(t)+C_{ext}(t)  
\label{langevinrot}
\end{equation}
where $\alpha$ is a friction coefficient and $C_{ext}$
denotes an external torque to which the system may also be submitted.
When no external torque is applied ($C_{ext}=0$), a
useful quantity that can be extracted from this equation is the power
spectral density $S$, defined as twice the Fourier transform of the
auto-correlation function $\langle \theta(t)\theta(t')\rangle$
($\langle\ldots\rangle$ denotes the statistical average over the noise). Using
the Wiener-Khintchine theorem for stationary processes, we get
\begin{equation}
S(\omega) = \frac{2q}{I^2(\omega^2-\omega_0^2)^2+\alpha^2\omega^2}
    \label{PSD}
\end{equation}
where $\omega_{0}=\sqrt{G/I}$ is the natural angular frequency of the
oscillator.

On the other hand, one can also focus on how the oscillator responds
to an external torque $C_{ext}(t)$. The quantity
containing this information is the susceptibility $\chi(t)$ (or linear
response function), defined as
\begin{equation}
\theta(t) = \int \!dt' \chi(t-t')C_{ext}(t')
\end{equation}
which implies that the external torque $C_{ext}(t)$
 should be small enough for this linear approximation to be valid.
From this definition and from the Langevin
equation~(\ref{langevinrot}), we see that in the Fourier
representation
\begin{equation}
    \chi(\omega) = \frac{\theta(\omega)}{C_{ext}(\omega)}
    = \frac{1}{I\left(\omega_0^2-\omega^2\right)+i\alpha \omega}
    \end{equation} The real and imaginary parts of this complex
    function can be defined as $\chi(\omega) =
    \chi'(\omega)-i\chi''(\omega)$, where in particular
\begin{equation}
    \chi''(\omega) = \frac{\alpha\omega}{I^2
    (\omega^2-\omega_{0}^2)^2+ \alpha^2\omega^2}\label{imchi}
    \end{equation} 
\noindent Comparing the power spectral density (\ref{PSD}) and the
imaginary part of the susceptibility (\ref{imchi}), we now notice that
these two very different concepts have similar expressions and are
related by the simple relation
\begin{equation}
    \frac{S(\omega)\omega}{\chi''(\omega)} = \frac{2q}{\alpha}
   \end{equation}

In a thermal system at equilibrium, using the equipartition of energy
principle, the parameter $q$ can be related to the bath temperature as
$q=2\alpha k_{B}T$, thus giving the celebrated Fluctuation-Dissipation
Theorem which states that 
\be
S(\omega)\omega/\chi''(\omega) = 4k_{B}T.
\label{fdt_langevin}
\ee

Since a vibrated granular medium is not at equilibrium,
there is no reason, in principle, to expect that a relation like
(\ref{fdt_langevin}) should hold for such a system. We can expect
that a granular medium can be found in quasi-stationary states but 
no ergodic principle can be invoked whatsoever.  Nevertheless if
a simple relation like (\ref{fdt_langevin}) were valid for a
granular medium, at least in some regime, this would be a strong hint
for the comprehension of the thermodynamic properties of such systems. 

The experiment we present was aimed precisely at the check of relation
(\ref{fdt_langevin}). The idea is to measure the noise power spectrum
and the susceptibility while the granular medium is externally driven
in the liquid-like state, and test whether there exists a
Fluctuation-Dissipation-like relation. If so, this will give us a
measure of a temperature-like parameter.  

\subsection{Experimental setup}

\begin{figure}[h]
        \begin{center}
\centerline {
\includegraphics[clip=true,width=4cm,keepaspectratio]{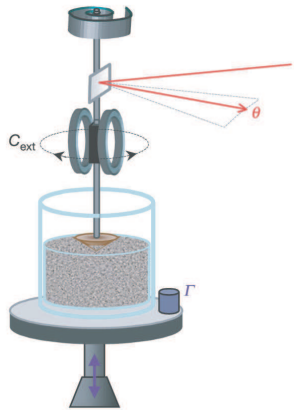}
}
\end{center}
\caption{Sketch of the torsion oscillator immersed into the ``granular
bath''.  The granular medium, composed of glass beads of diameter
$1.1\pm 0.05$~mm is placed in a cylindrical container of height 60~mm
and diameter 94~mm. An accelerometer measures the intensity of the
external perturbations, $\Gamma$.}
\label{experiment}
\end{figure}

We use the following experimental setup~\cite{dannanature}: a thin
torsion oscillator is immersed at some depth into a granular medium
made of millimeter-size glass beads, as shown in
Fig.~\ref{experiment} (note the
analogy with the situation described in~\cite{uhlenbeck}, for a system
at thermal equilibrium).
The beads are placed into a cylindrical container which is
continuously vibrated vertically, with a high-frequency filtered white
noise (cut off above 900~Hz and below 300~Hz in the experiments
described). We use this vibration mode to ensure a homogeneous
agitation and avoid undesired effects such as pattern formation and
convection rolls. Note that this type of white noise vibrations is not
used in order to provide a random torque with white noise spectrum to
the oscillator: actually, its motion is observed in a much lower
frequency range (10-50 Hz) than the vibrations applied.

The vibration intensity is determined by an accelerometer fixed on the
container, which measures the parameter $\Gamma$, defined as
$\Gamma^2=\int \!\!A(f)\,df$, where $A(f)$ is the acceleration
spectrum, normalized to the acceleration of gravity, and the
integration is taken in the frequency range of about 1~Hz to 10~kHz.
For sinusoidal vibrations, $\Gamma=1$ is the ``fluidization''
threshold, above which a single grain starts to fly. Here, we
typically use vibration intensities between $\Gamma=1$ and $15$.

The oscillator angular position $\theta$ is detected optically (see
Fig.~\ref{experiment}). For susceptibility measurements, two external
coils and a permanent magnet fixed on the oscillator axis allow to
apply a sinusoidal torque
$C_{ext}(t)=C_{e}\sin(\omega t)$.

\subsection{Results and discussion}

\begin{figure}[h]
\begin{center}
\centerline {\includegraphics[clip=true,width=12cm,
keepaspectratio]{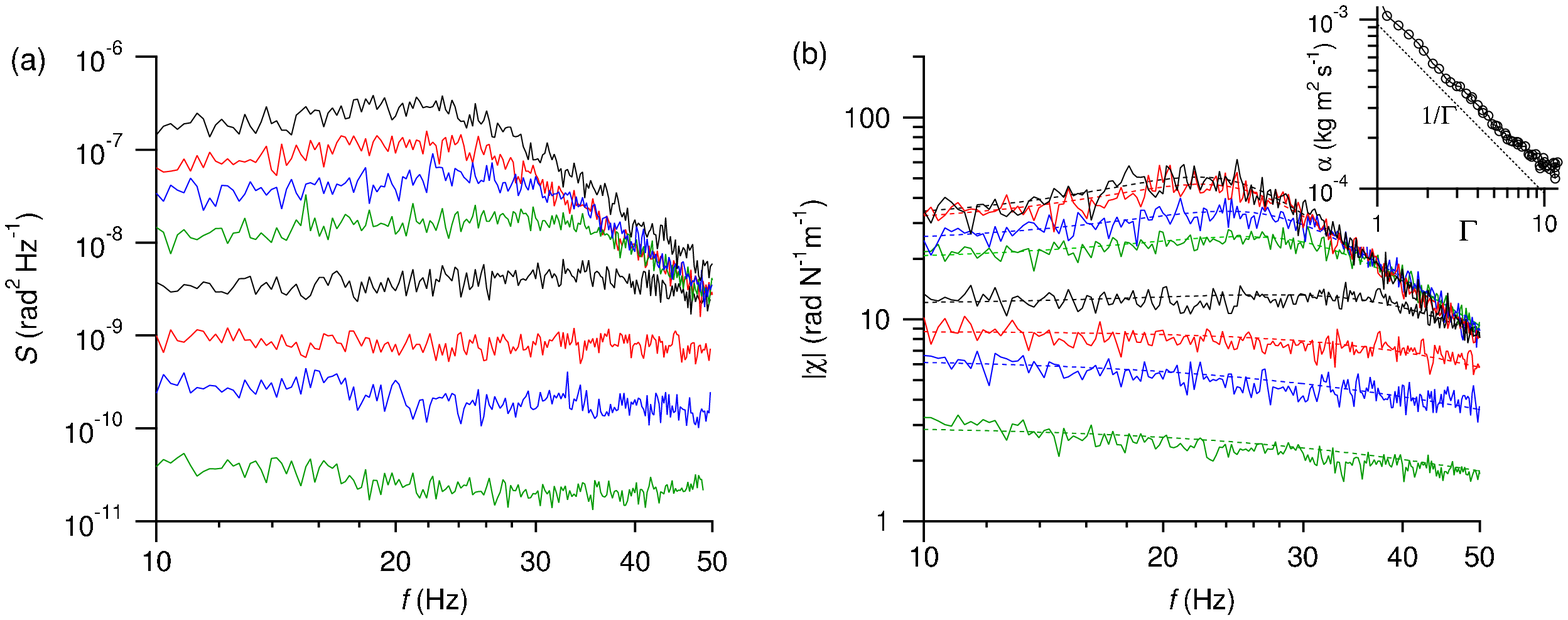}}
\end{center}
\caption{Power spectral density (a) and modulus of the complex
susceptibility (b) versus the frequency, $f=\omega/2\pi$, for
different vibration intensities $\Gamma$: from top to bottom, 11.6,
10.0, 7.3, 5.4, 3.7, 2.2, 1.5, 1.0. Results obtained with a conical
probe of apex angle 120б, covered with a single layer of glass beads.
The immersion depth is about 11~mm.  In~(b), each curve is fitted
(dashed line) with its corresponding equilibrium expression (see text
for details). The moment of inertia of the oscillator is $I=1.5\times
10^{-6}$~kg$\cdot\textnormal{m}^2$, and the applied torque amplitude
is $C_{e}=3.2\times 10^{-5}$~N$\cdot$m. Inset: Friction
coefficient~$\alpha$, obtained from the fit to the curves
$|\chi(\omega)|$, versus~$\Gamma$. The dotted line is a power law
$\alpha\propto 1/\Gamma$.}
\label{Schi}
\end{figure}

The analysis of angular deflection time-series $\theta(t)$ in the
free mode ($C_{ext}=0$) provides the noise power spectral
density $S(\omega)$, shown in Fig.~\ref{Schi}a for different values of
$\Gamma$. Then, with the oscillator in forced mode (with an externally
applied torque $C_{ext}(t)$), we measure the complex
susceptibility $\chi(\omega)$, whose modulus $|\chi(\omega)|$ is shown
in Fig.~\ref{Schi}b. The amplitude of the external torque is small
enough to be in the regime of linear response.

Fitting these curves $|\chi(\omega)|$ with the standard expression for
the damped oscillator $|\chi(\omega)| =
[I^2\left(\omega_0^2-\omega^2\right)^2+ \alpha^2\omega^2]^{-1/2}$
shows a good agreement, thus supporting the idea that the Langevin
equation of motion (\ref{langevinrot}) is a pertinent description of
the oscillator linear response.  This allows us to extract a granular
friction coefficient~$\alpha$, or a granular viscosity
$\mu\propto\alpha$, found to be inversely proportional to the
vibration intensity: $\alpha\propto 1/\Gamma$ (see inset of
Fig.~\ref{Schi}b).

\begin{figure}[h]
    \begin{center} \centerline
{\includegraphics[clip=true,width=10cm,keepaspectratio]{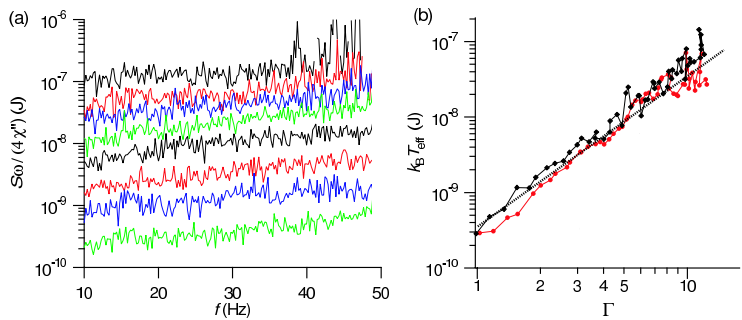}}
\end{center}
\caption{(a) Fluctuation-Dissipation ratios
$S(\omega)\omega/(4\chi''(\omega))$ versus the frequency
$f=\omega/2\pi$ for different vibration intensities (as in
Fig.~\ref{Schi}) (b) Effective temperature versus vibration
intensity~$\Gamma$, as obtained from Fluctuation-Dissipation levels in
(a) averaged between 10 and 50~Hz (black symbols), and from experiment
using a conical probe with triple moment of inertia (red symbols). A
power law fitted to the data gives $T_{eff}\propto
\Gamma^{p}$, with $p=2.1$. The dashed line has equation $k_{B}
T_{eff} = 3.5\times10^{-10} \,\Gamma^2$.}
\label{FD}
\end{figure}

We can also calculate the Fluctuation-Dissipation ratios
$S(\omega)\omega/(4\chi''(\omega))$, which are shown in
Fig.~\ref{FD}a. These ratios, even though not constant, are
surprisingly `flat', especially compared to what has been measured in
other non-equilibrium thermal systems, such as in
glycerol~\cite{grigera:1999} and laponite~\cite{bellon}.

This reveals that the high-frequency driven agitation of the granular
medium acts on the oscillator as a source of random torque with white
spectrum, at least in the 10-50~Hz range under consideration. Energy
is thus injected at high frequency, and spreads into a low frequency
white spectrum.

Since those ratios do not exhibit a strong frequency dependence, this
provides support for the existence of a Fluctuation-Dissipation
relation in off-equilibrium driven granular steady states. This relation
can thus be used to define an effective
temperature,~$T_{eff}$.  Figure~\ref{FD}b shows the
averaged Fluctuation-Dissipation ratio levels, that is,
$k_{B}T_{eff}$, versus~$\Gamma$ (black curve).  Fitting
to a power law yields $k_{B}T_{eff}\propto \Gamma^p$ with
$p$ close to 2.
This dependence suggests that, due to the complex
dissipation processes between the grains, a fixed fraction of the
energy input (vibrations) is effectively available as granular kinetic
energy and is ``sensed'' by the oscillator.

In fact, we notice that the order of magnitude of the thermal energy
$k_{B}T_{eff}$, as measured here, is consistent with
realistic values of the mean kinetic energy per particle, as measured
by grain-tracking methods~\cite{dixon, yang}.  Thus, the effective
temperature $T_{eff}$ measured seems to be related to the
granular temperature, as usually defined in granular
gases. This is particularly interesting in view
of the recent numerical experiments relating the granular temperature with 
a dynamical temperature obtained through FDT-like
measures in granular gases (see section \ref{section:gas} and
\cite{Puglisi:2002,Barrat:2004}).

These results indicate that the use of macroscopic quantities, such as
temperature-like or viscosity-like parameters, to describe the
behavior of externally driven granular media is in first approximation
possible.  We can notice that, as shown in Fig.~\ref{FD}b, the
temperature parameter obtained does not seem to depend on the moment
of inertia of the probe used.  On the other hand, we have also
indications (not discussed in detail here) that several complications,
related to the discrete and inhomogeneous nature of granular media and
to the role of gravity, have to be taken into account.  These
mechanisms are triggered on as soon as the probe-granular medium
interactions take place on length scales that are comparable to the
grain dimensions, for instance when using non-smooth probe sections
exhibiting features with length scales of the same order as the grain
diameter, or close to the surface\ldots The systematical study of
these geometrical and grain-level effects will give us the possibility
to establish empirical laws that may be used to predict granular
behavior in practical situations, and possibly in simulations.

Let us note that the temperature measured here is not a priori the
Edwards temperature. However, by decreasing the external vibrations we
observe evidence of a glassy behavior~\cite{dannaprl:2001}.  In
particular, the study of the power spectrum (hence of the diffusivity)
reveals that for very weak external vibrations the granular medium
exhibits a critical slowdown from the fluid state to a glassy state,
where - as for supercooled liquids - the diffusivity approaches
zero. In thermal systems undergoing a glass transition, the definition
of a temperature is an open issue, and there are suggestions proposing
that a ``configurational temperature'', sharing similarities with the
Edwards temperature, can be introduced. Thus, in the limit of very
small external vibrations, it is possible that we are confronted with
a similar problem in the granular media.

\section{Granular gases} 
\label{section:gas}

In this section the particular case of highly fluidized granular materials is
taken into account. When a box full of grains is strongly shaken and the
volume available is large enough, the assembly of grains behaves in a way
similar to standard molecular gases. In the literature a large series of
experiments has led to the definition of ``granular gases'', i.e. granular
material in a gaseous state~\cite{po1,po2}. Because of the analogy with usual
gases, the term "granular temperature" has naturally been defined as the
average local kinetic energy per particle: this definition indeed coincides
with the thermodynamic equilibrium temperature in the limit of 
elastic collision between particles.

In experiments, it is possible to study
kinetic as well as hydrodynamic observables and compare them with the
results of statistical mechanics and hydrodynamics of molecular
gases. The observed differences may be explained as a consequence of
loss of energy in collisions between grains.  Energy dissipation
during a collision is due to irreversible transfer of energy from
macroscopic energy to internal degrees of freedom and eventually to
heat. Therefore it can be said that {\em inelasticity} is the main
ingredient in the description of a granular material in the dilute
regime, while frustration and excluded volume effects are
negligible. We will show that, while inelastic collisions produce
large deviations with respect to usual thermodynamics and
hydrodynamics, linear response theory and Fluctuation-Dissipation
relations are still valid~\cite{Puglisi:2002,Barrat:2004,Garzo:2004},
provided that the ``equilibrium temperature'' is replaced by the
granular temperature of each component of the gas.

\subsection{Fluctuation Dissipation relations for diluted granular systems}

We have studied Fluctuation-Dissipation relations using two sets of
independent measurements, i.e. two choices of the pair
response-correlation.

{\em Recipe I}: The first one consists in the classical measure of
mobility and diffusivity. The mean-square displacement
\begin{equation}
B(t,t') = \frac{1}{N}\sum_{j=1}^{N} 
\langle |{\bm r_j}(t) - {\bm r_j}(t')|^2 \rangle  \ ,
\end{equation}
behaves asymptotically as $\sim 4 D(t-t')$. The mobility of a tracer
particle can be measured by applying a constant and small drag force
$\boldsymbol{\xi} =\xi {\bm e_x}$ to a given particle, labeled $0$,
for times $t > t'$ (linearity of the response has been checked by
changing the amplitude of the perturbation).  The perturbed particle
will reach at large times a constant velocity $\mu$, related to the
response by
\begin{equation}
\chi (t,t') = \frac{1}{\xi} \langle ({\bm r_0 (t)} - {\bm r_0
(t')})\cdot {\bm e_x} \rangle \approx \mu t \ \ \mbox{at large\ t}.
\end{equation}
At {\em equilibrium} Green-Kubo relations (Einstein relation in this
case) predict $\mu=\beta D$, $T_b=1/\beta$ being the equilibrium
temperature, so that $\chi (t,t') = \frac{\beta}{4} B$.

{\em Recipe II}: Another totally independent way of checking FD
relations is the following: once a steady-state has been reached, the
system is perturbed impulsively at a given time $t_0$ by a
non-conservative force applied (non-uniformly) on every particle. The
response is then monitored at later times. The force acting on
particle $i$ is
\begin{equation} \label{forcing}
\mathbf{F}(\mathbf{r}_i,t)=\gamma_i \boldsymbol{\xi}(\mathbf{r}_i,t)
\end{equation}
with the properties $\boldsymbol{\nabla} \times \boldsymbol{\xi} \neq
0$, $\boldsymbol{\nabla} \cdot \boldsymbol{\xi} = 0$, where $\gamma_i$
is a particle dependent variable with randomly assigned $\pm 1$
values. A simple case is realized by a transverse perturbation
$\boldsymbol{\xi}(\mathbf{r},t)= (0, \xi \cos (k_x x) \delta(t) )$,
where $k_x$ is compatible with the periodic boundary conditions,
i.e. $k_x=2 \pi n_k /L_x$ with $n_k$ integer and $L_x$ the linear
horizontal box size.  The staggered response function $R$ (i.e. the
current induced at $t$ by the perturbation at $t_0$), and the
conjugated correlation $C$,
\begin{eqnarray} 
R(t,t_0) & = & \frac{1}{\xi}\langle \sum_i \gamma_i \dot{y}_i(t)
\cos(k_x x_i(t)) \rangle \ ,\nonumber \\ 
C(t,t_0) & = &\langle \sum_i
\dot{y}_i(t) \dot{y}_i(t_0) \cos \{k_x[x_i(t)-x_i(t_0)] \} \rangle
\nonumber
\end{eqnarray}
are related, {\em at equilibrium}, by the FD relation $R(t,t_0) =
\frac{\beta}{2} C(t,t_0)$, $T_b=1/\beta$ being the equilibrium
temperature.

We have applied both recipes to a model of pure granular gas as well
as to a model of granular binary mixture. We have also applied recipe
I to a much simplified model of pure granular gas (inelastic Maxwell
model) where analytical calculations of diffusion and mobility can be
straightforwardly obtained.

\subsection{Description of the models}

The simplest model of granular material in two dimensions is the hard
disks gas with inelastic collisions.  To counterbalance the loss of
energy we enforce the stochastic forcing model, i.e. particles receive
random acceleration as if they were in contact with a ``heat
bath''. Moreover, we may (or may not) provide a viscous drag to each
particle. Viscosity has the role, in this model, of a regularizing
force and allows for a better definition of elastic limit. It can be
thought as the result of friction with external walls or with a fluid
the gas is immersed in. However we will show that viscosity is not
essential in this study and identical results can be obtained without
it.

We consider a volume $V$ in dimension $d=2$ containing $N=N_1+N_2$ inelastic
hard disks, $N_1$ and $N_2$ being the
number of particles in component $1$ and $2$ of the mixture,
respectively. The disks have diameters $\sigma$ (identical for the
two species) and masses $m_{s_i}$ (where $1 \leq i \leq N$ and $s_i$
is the species index, $1$ or $2$, of particle $i$).  In a collision
between spheres $i$ and $j$, characterized by the inelasticity
parameter called coefficient of normal restitution $\alpha_{s_is_j}$,
the pre-collisional velocity of particle $i$, ${\bm v}_i$, is
transformed into the post-collisional velocity ${\bm v}'_i$ such that
\begin{equation}
\bm{v}_i' \, =\,  \bm{v}_i - 
\frac{m_{s_j}}{m_{s_i}+m_{s_j}} (1+\alpha_{s_is_j})
(\widehat{\bm{\sigma}}\cdot \bm{v}_{ij})\widehat{\bm{\sigma}} 
\label{eq:coll1}
\end{equation}
where $\bm v_{ij}=\bm v_i-\bm v_j$ and $\widehat{\bm\sigma}$ is the center
to center unit vector from particle $i$ to $j$
($\alpha_{s_is_j}=\alpha_{s_js_i}$ so that the total linear momentum $m_i
\bm v_i+m_j\bm v_j$ is conserved).

In between collisions, the particles are submitted to a random force in
the form of an uncorrelated white noise (e.g. Gaussian) with
the possible addition of a viscous term. The equation of motion for
a particle is then
\begin{equation}
m_i \frac{d {\bm v}_i}{dt} = {\bm F}_i + {\bm R}_i - \zeta_{s_i} {\bm v}_i
\end{equation}
where ${\bm F}_i$ is the force due to inelastic collisions,
$\zeta_{s_i}$ is the viscosity coefficient and $\langle R_{i\alpha}(t)
R_{j\beta}(t') \rangle = \xi_{s_i}^2 \delta_{ij}\delta_{\alpha\beta}
\delta(t-t')$, where Greek indexes refer to Cartesian coordinates.
The granular temperature of species $s$ is given by its mean kinetic
energy $T_{s} = m_{s} \langle v^2\rangle_{s}/d$ where $\langle
...\rangle_s$ is an average restricted only to particles of species
$s$.

{\em Pure system}: when $m_1=m_2=m$, $\zeta_1=\zeta_2=\zeta$,
$\xi_1=\xi_2=\xi$ and $\alpha_{11}=\alpha_{12}=\alpha_{22}=\alpha$, the
gas is monodisperse. This model has been extensively
studied~\cite{Williams,Puglisi,Twan,Twanetal,Montanero,Cafiero,Moon,nicodemi_vulpiani}. When
$\zeta \neq 0$ a ``bath temperature'' can be defined as
$T_b=\xi^2/2\zeta$. This corresponds to the temperature of a gas
obeying equation~(\ref{eq:coll1}) with elastic collisions or without
collisions. The same temperature can be observed if the viscosity is
very high, i.e. when $\zeta \gg 1/\tau_c$ where $\tau_c$ is the mean
free time between collisions. Here we recall that when $\alpha<1$ the
gas still attains a stationary regime, but its granular temperature is
smaller than $T_b$ and therefore the system is out of
equilibrium. Moreover, the statistical properties of the gas are
different from those of an elastic gas in contact with a thermal bath:
mainly the velocity distribution is non-Gaussian with enhanced
high-energy tails. The system is usually studied by means of molecular
dynamics (with hard core interactions) or by means of numerical
solutions of the associated Boltzmann equation. As we are interested
in the dilute case, where Molecular Chaos is at work, we follow this
second recipe, implementing the so-called Direct Simulation Monte
Carlo~\cite{bird} (DSMC) which is a numerical scheme that solves the
Boltzmann equation for homogeneous or (spatially) non-homogeneous
systems.

{\em Binary mixture}: in the general binary mixture case, simulations
as well as experiments and analytical calculations have shown that
energy equipartition is broken, i.e. $T_1 \neq T_2$.  At the level of
Boltzmann kinetic equation, the temperature ratio of a binary granular
mixture subject to stochastic driving of the form given above has been
obtained in~\cite{Equipart} for the case $\zeta_{s_i}=0$ and
in~\cite{Pagnani} for $\zeta_{s_i} \neq 0$. In the case $\zeta_{s_i}
\neq 0$ a bath temperature can still be defined as
$T_b=\xi_{s_i}^2/2\zeta_{s_i}$. Note that in general $\xi_{s_i}$ and
$\zeta_{s_i}$ depend on $m_i$ and the correct elastic limit
(i.e. equipartition) is recovered if and only if $T_b$ does not depend
upon $m_i$. In~\cite{Pagnani} it has been shown that a model with
$\xi_{s_i} \propto \sqrt{m_{s_i}}$ and $\zeta_{s_i} \propto m_{s_i}$
fairly reproduces experimental results for the temperature ratio
$T_1/T_2$ measured in a gas of grains in a vertically vibrated box. It
is also known that equipartition is not recovered even in the
so-called tracer limit~\cite{Martin}, i.e. in the case $N_2=1$ and
$N_1 \ll 1$. For binary mixtures we have implemented both molecular
dynamics and DSMC, in order to study possible differences.


{\em Inelastic Maxwell Model}: in analogy with elastic gases, some simplified model can be
introduced. For instance, it has recently been
proposed~\cite{bennaim-krapivsky} the inelastic analogue of a Maxwell
gas~\cite{ernst-review}.  An elastic Maxwell gas is made by particles
interacting through a special repulsive long range potential. The
interest of such an interaction is that the corresponding collision
frequency results strongly simplified: while for hard disks the
collision frequency is proportional to the relative impact velocity
$g$, for particles of a Maxwell gas it becomes independent of
$g$. So the resulting Boltzmann equation takes a much simpler
convolutive expression. The stochastic model of inelastic Maxwell
gases is directly defined introducing a normal restitution coefficient
smaller than one in the kinetic equation of an elastic Maxwell gas. As
in the elastic case, the interest for such operation is the
achievement of interesting exact results. 

In the simplest case, i.e. in one dimension, the Boltzmann equation of
an inelastic Maxwell gas reads $\partial_t
P(v,t) = \beta \int du\, P(u,t)P\left(\beta v\!+\!(1-\beta)
u,t\right)-\!P(v,t)$, where $\beta=2/(1+\alpha)$ and $\alpha$ is the
restitution coefficient.  At odds with the elastic case, its only
stationary solution is the degenerate delta function, representing a
system of particles at rest. Starting with a finite energy, the
cooling of the gas has an easily computable rate, since the average
squared velocity per particle decays exponentially $v_0(t)^2=\int v^2
P(v,t) dv = v_0(0)^2 \exp(-\lambda t)$ with $\lambda= (1-\alpha^2)/2$.

One of the interesting feature of this model, is that it admits a
scaling solution of the form $P(v,t)=\frac 1{v_0(t)} f(v/v_0)$,
although the scaling function $f$ cannot be recovered through a simple
analysis of the velocity momenta~\cite{bennaim-krapivsky}.  In fact,
it can be shown that the scaling function that solves the equation,
$f(c)=\frac 2\pi \frac 1{\left[1+c^2\right]^2}$, is a distribution
whose moments higher than the second diverge.  Note that, while the
rate of dissipation depends on the restitution coefficient, the
scaling solution does not.  This is peculiar of the one dimensional
case. In higher dimension~\cite{ernst-bennaim-2d} the algebraic tails
of the solution are still present, but the exponent depends on the
restitution coefficient (the tail is narrower and narrower reducing
the inelasticity and the distribution becomes Gaussian in the elastic
limit). The observation of scaling solutions for Maxwell gases is not
new. However, for the elastic case, the solutions are not relevant,
while in the inelastic case an initial distribution (e.g. an
exponential or uniform) rapidly converges to the scaling solution. If
we consider an initial distribution with a finite energy per particle
and we define $f(c,t) \equiv v_0(t) P(c v_0,t)$, then the equation for
the inelastic Maxwell model can be recasted into
\begin{equation}
\label{stationary-gaussianbath}
\partial_t f(c,t) + \lambda \partial_c \left( c f(c,t)\right)  = \beta \int ds f(s,t)
f(\beta c + (1-\beta)s) - f(c)
\end{equation}
This equation can be read as the Boltzmann equation of a new system,
which is an inelastic Maxwell gas submitted to a special driving bath
(it is often called ``Gaussian thermostat'') given by the term
proportional to $\partial_c
(cf(c,t))$~\cite{santos-transport-coeff}. Such a gas performs a
stationary dynamics where the energy lost by inelastic collision is
compensated by a negative viscosity term, which pushes the particles
with a force proportional to their velocity.  As we shall see in the
following, it is possible to straightforwardly compute the mobility
and the diffusion coefficient for a simple stochastic model governed
by equation~(\ref{stationary-gaussianbath}). This allows to explicitly
check the validity of the Einstein relation for such a stationary
non-equilibrium dynamics.

\subsection{Results}

\begin{figure}[ht]
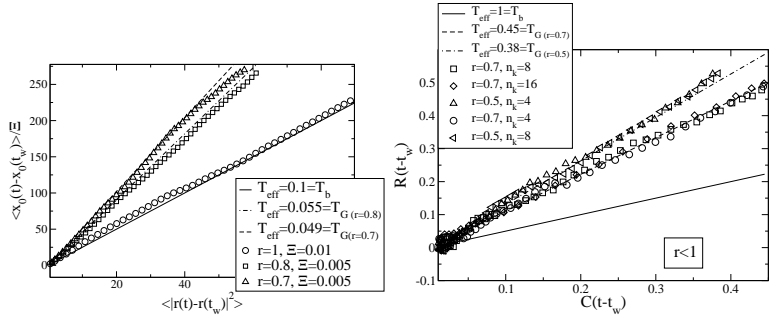

\centerline {\includegraphics[clip=true,width=5cm,
keepaspectratio]{diff_1.eps}
\includegraphics[clip=true,width=5cm,
keepaspectratio]{cicc_inel.eps}}
\caption{\label{fig:pure} {\bf Pure system}. Left: parametric plot of $\chi(t,t_w)$
vs. $B(t,t_w)$ for the numerical experiment with recipe $I$ (constant
force applied on one particle) with $r=1$, $r=0.8$ and $r=0.7$, with heating
bath, and for different choices of the intensity $\Xi$ of the
perturbation, using $T_b=0.1$ ad $\tau_b=10$, $N=500$, $\tau_c=1$,
$t_w=100$. The results are obtained by averaging over $10000$
realizations. Right: parametric plot of $R(t-t_w)$
vs. $C(t-t_w)$ for the numerical experiment with recipe $II$ (impulsive
shear perturbation) with $r<1$, with heating bath, and for different
choices of the wave number $n_k$ of the perturbation. $T_b=1$ and
$\tau_b=10$, $N=500$, $\tau_c=1$, $\Xi=0.01$, $n_k=8$, with averages
over $10000$ realizations, using $t_w=100$.}
\end{figure}

 {\em Pure systems}: in Fig.~\ref{fig:pure}, left frame, a parametric plot of mobility versus
diffusion is displayed for several choices of parameters, showing the
linear behavior analogous to Green-Kubo formulas (Einstein relation in
this case). The same linear behavior is recovered plotting response
versus perturbations in the case of an impulsive shear perturbation
(recipe II experiment, see Fig.~\ref{fig:pure}, right frame). From the slope
$s$ of the observed linear behavior in the response-perturbation
graph, one can get the effective temperature $T_{eff}=1/(4s)$. We
always find $T_{eff}=T_g$, with $T_g \leq T_b$.

\begin{figure}[htb]
\centerline{ \psfig{figure=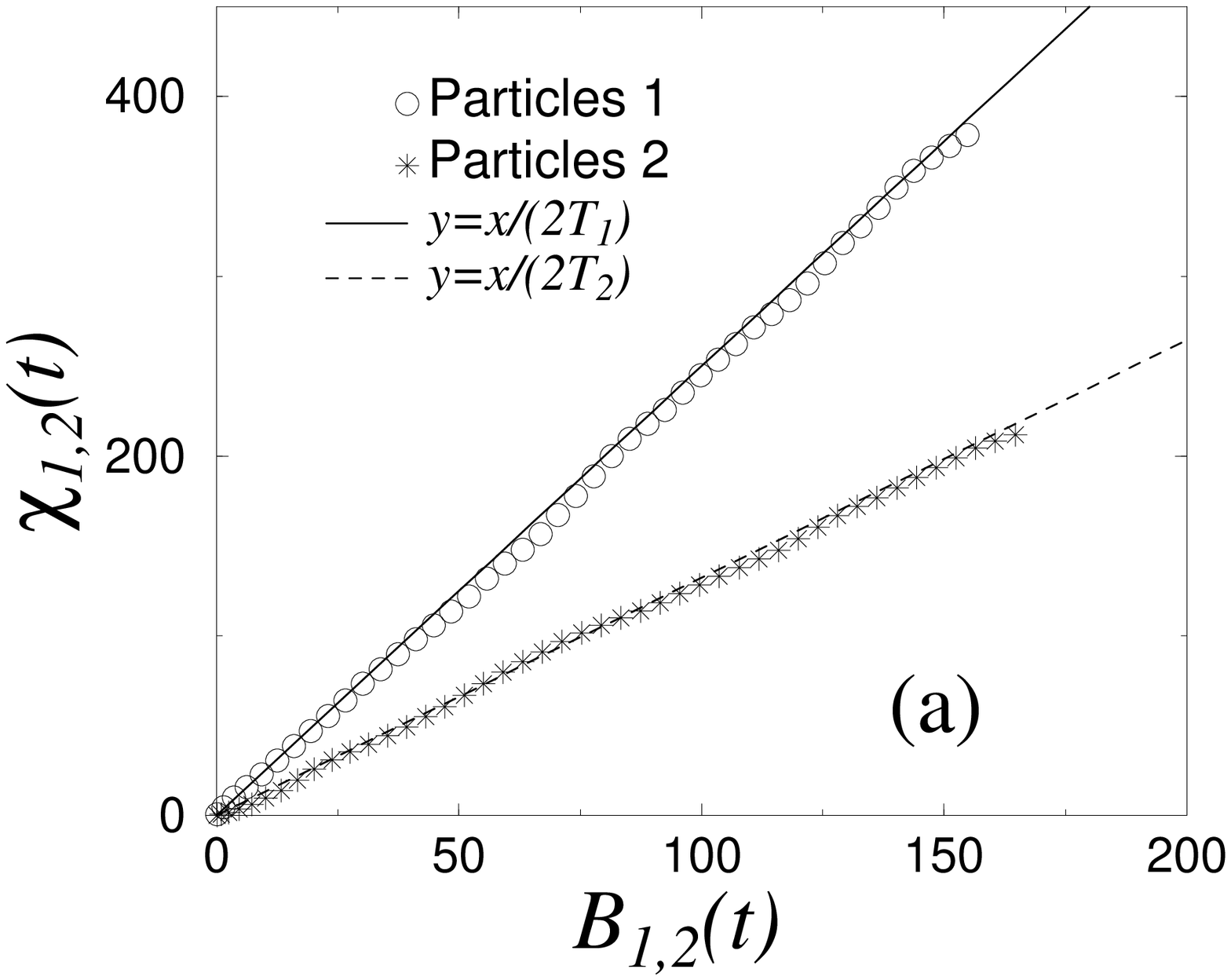,width=4cm,angle=0}
        \psfig{figure=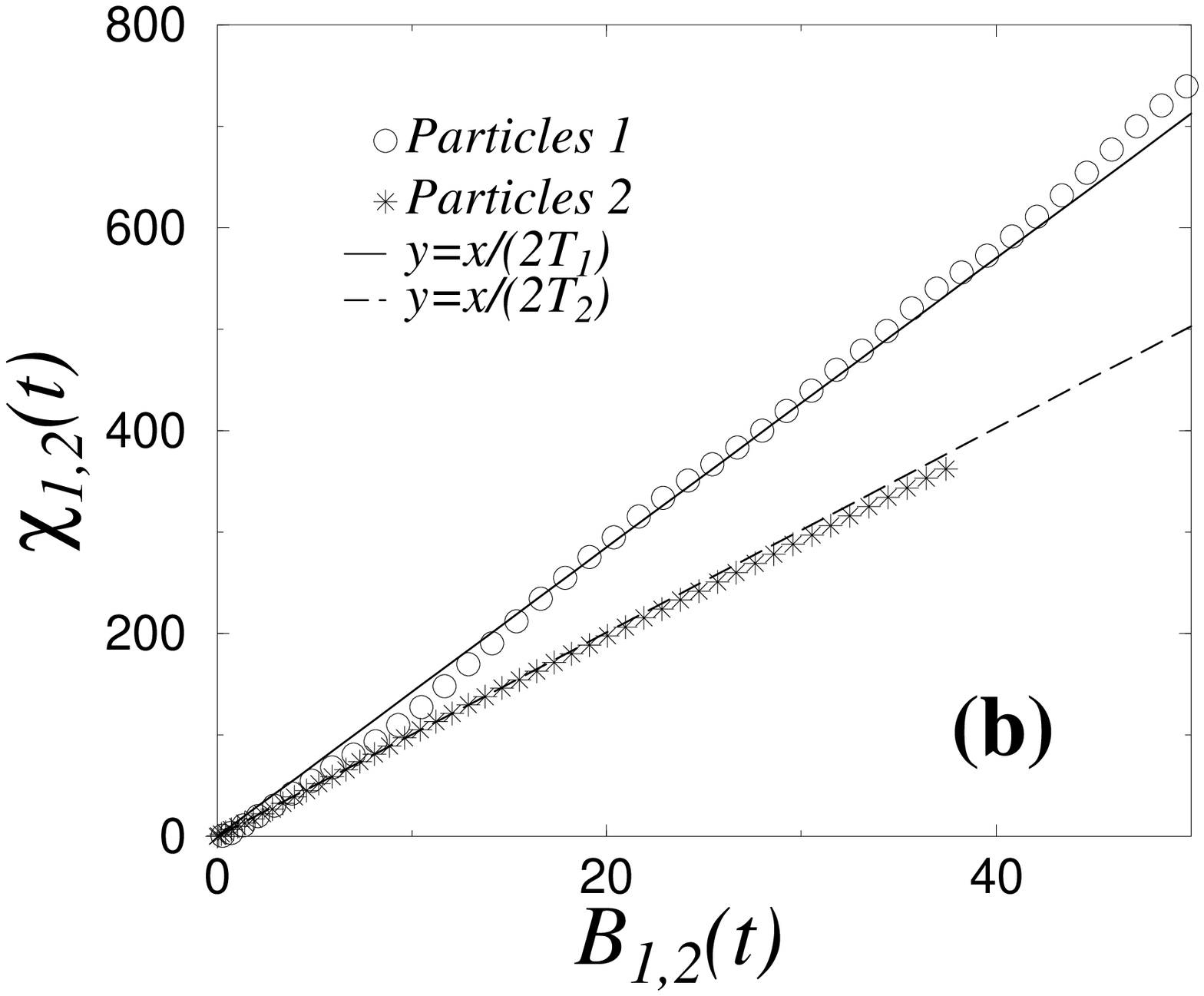,width=4cm,angle=0} }
\caption{Binary mixture, homogeneous DSMC: mobility vs. mean-square
displacement; left: $\alpha_{11}=0.3$, $\alpha_{12}=0.5$,
$\alpha_{22}=0.7$, $m_2=3m_1$, $T_1\approx 0.2$, $T_2\approx 0.38$;
right: $\alpha_{11}=\alpha_{12}=\alpha_{22}=0.9$, $m_2=5m_1$,
$T_1\approx 0.035$, $T_2\approx 0.05$.  Symbols are numerical data,
lines have slope $1/(2T_1)$ and $1/(2T_2)$.  }
\label{fig:hom_kicks}
\end{figure}

{\em Binary mixtures}: in the case of binary mixtures, linearity or
response-perturbation relations is again verified, in the
mobility-diffusion experiment as well as in the current-shear
perturbation experiment~\cite{Barrat:2004}. Here we review just the
former results, i.e. those for the Einstein relation. By successively
using as test particle one particle of each species, one obtains the
two responses $\chi_1$ and $\chi_2$, and thus the mobilities $\mu_1$
and $\mu_2$. Two independent Einstein relations ($\mu_i= 2 D_i /T_i $)
are verified, by plotting $\chi_i$ vs. $B_i$. In
Fig.~\ref{fig:hom_kicks} we show, as an example, the check of the
validity of the Green-Kubo relations using DSMC in spatially homogeneous
regime. All the experiments, performed varying the restitution
coefficients and the masses of the two components, and with different
models and algorithms (homogeneous and inhomogeneous, DSMC and MD)
showed identical results, i.e the linearity of the response-perturbation
relation with the effective temperature equal to the granular temperature
of the perturbed species.

\begin{figure}[tb]
\centerline{ \psfig{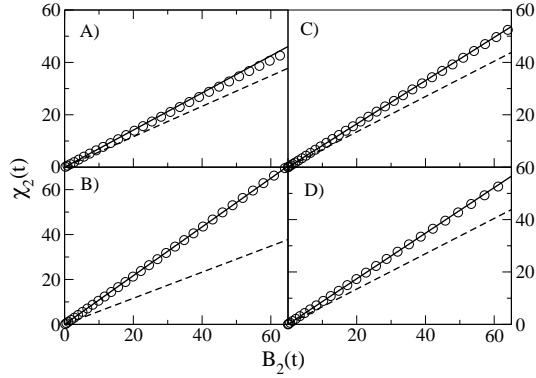} }
\caption{Binary mixture, homogeeous DSMC: mobility Vs. diffusion of a single particle of mass
$m_{tracer}$ in contact with $N=500$ particles of mass $m$, immersed
in a heat bath (i.e. random kicks plus viscosity).  We use the
following conventions: $\alpha_{tracer}=\alpha_{12}$ and
$\alpha=\alpha_{11}$. Only in case A) the tracer is also in contact
with the external driving heat bath.  {\bf A)} $m_{tracer}=m$,
$\alpha=0.9$, $\alpha_{tracer}=0.4$, $T_g=0.86$, $T_g^{tracer}=0.70$;
{\bf B)} $m_{tracer}=m$, $\alpha=0.9$, $\alpha_{tracer}=0.4$,
$T_g=0.86$, $T_g^{tracer}=0.46$; {\bf C)} $m_{tracer}=7m$,
$\alpha=\alpha_{tracer}=0.7$, $T_g=0.74$, $T_g^{tracer}=0.60$; {\bf
D)} $m_{tracer}=4m$, $\alpha=\alpha_{tracer}=0.7$, $T_g=0.74$,
$T_g^{tracer}=0.57$.  The solid line has slope $T_g^{tracer}$, the
dashed line has slope $T_g$.  }
\label{fig:tracer}
\end{figure}

In Fig.~\ref{fig:tracer} an even more striking result is portrayed:
the mobility-diffusion parametric graph is shown in the case of a
single tracer with different properties with respect to a bulk gas
($N_1=500$, $N_2=1$). In this case the tracer does not perturb
significantly the bulk. However the temperature of the tracer is quite
different from the bath temperature as well as from the gas
temperature~\cite{Martin}. Again, the effective temperature of the
tracer correspond to {\it its} temperature and not to the temperature
of the bath or of the bulk. This is equivalent to say that a
non-perturbing thermometer, used to measure temperature of a granular
gas through Fluctuation-Response relations, would measure its own
temperature and not the bulk temperature.

{\em Exact solution of the inelastic Maxwell model}:

Here we present a simple calculation on the inelastic one-dimensional
Maxwell gas driven by a Gaussian bath where the (non-equilibrium)
Einstein relation suggested by numerical simulation holds.

Let us consider a gas of particles performing binary inelastic collision
with a constant collision rate (i.e. independent of the
precollisional relative velocity) and submitted to a negative
viscosity force, which drives the gas in a stationary state. The
variation of the velocity of a particle after a time $\Delta t$ is
given by:
\begin{equation}
v(t+\Delta t)-v(t) = \left\{
\begin{array}{cc} 
\lambda v(t) \Delta t & \mbox{with prob. $1-\Delta t$}\\
-\frac{1+\alpha}2 (v(t)-u) +  \lambda v(t) \Delta t &  \mbox{with
  prob. $\Delta t$}
\end{array}
\right.
\label{particle-motion}
\end{equation}
where $u$ is the velocity of a generic colliding particle,
distributed as $f(u)$.
The autocorrelation function $A(t_1,t_2)=\overline{v(t_1) v(t_2)}$
(with $t_2>t_1$) can be computed~\cite{bennaim-autocorrelation} using ~(\ref{particle-motion}),
\[
\frac d{dt_2} A(t_1,t_2)=\lim_{\Delta t \rightarrow  0}
\overline{v(t_1)\frac{v(t_2+\Delta t)-v(t_2)}{\Delta t}}=\left[\lambda-\frac{1+\alpha}2\right] A(t_1,t_2),
\]
Since we are considering the stationary case, which is obtained with a
negative viscous force exactly balancing the dissipation,
i.e. $\lambda=(1-\alpha^2)/4$, this gives:
\[
A(t_1,t_2)=T_g \exp{\left[-\left(\frac{1+\alpha}2\right)^2(t_2-t_1)\right]}.
\]
The diffusion coefficient can be computed via the autocorrelation
function, obtaining:
\begin{equation}
\label{diffcoeff_maxwell}
D=\lim_{\tau\rightarrow \infty} \int_0^\tau d\tau' A(t_1,t_1+\tau') = \frac
{4 T_g}{(1+\alpha)^2}.
\end{equation}
Now, to compute the mobility, we have to apply a small constant force
$F$ to a particle, which modifies equation~(\ref{particle-motion}) so
that $\overline {\frac{d v(t)}{dt}}=-\frac{(1+\alpha)^2}4
\overline{v(t)} + F.$ The asymptotic velocity of the tracer yields $4
F/(1+\alpha)^2$.  This means that the mobility is
\[
\mu=\frac 4{(1+\alpha)^2}=\frac D{T_g}
\]
and the (non-equilibrium) Einstein relation holds.

\subsection{Summing up results about granular gases}

Approximated analytical results concerning Fluctuation-Dissipation
relations in granular gases have been obtained for the case of
mobility and diffusion of a tracer particle in homogeneous cooling
granular gases (i.e. without external driving) in~\cite{fd:cooling}
and in driven granular gases in~\cite{Garzo:2004}. In the cooling
case, two main ingredients spoil the usual Green-Kubo formula and lead
to a strong reformulation of Fluctuation-Response relations: they are
the strong non-Gaussian behavior of velocity distribution (in
Homogeneous Cooling State, HCS, the velocity p.d.f. has exponential
tails), and the non-stationarity due to the thermal cooling of the
gas. In the HCS, therefore, the Green-Kubo formula must be replaced by
a non-linear formula which takes into account these two strong
non-equilibrium effects. On the other side, the study of the tracer
kinetic equation (Boltzmann-Lorentz equation) for the driven case leads
to the conclusion that the only source of deviation from linearity in
the Fluctuation-Response graph may be the velocity non-Gaussian
behavior. However this is never very pronounced in granular gas driven
by a homogeneous source (e.g. when grains are on a table which is
vertically vibrated), so that deviations from linearity are
negligible. The replacement of $T_b$ by $T_g$, comes quite naturally
in the calculations. On the other side we have shown that in the
inelastic Maxwell model, in 1d, where the asymptotic velocity
p.d.f. is known to have power law tails, the FD relations is recovered
thanks to fortuitous balance of different terms.

The general lesson learnt from simulations and analytical calculation,
in the case of stationary dilute granular gas, is that FD relations
are difficult to be violated when ergodicity is at work with mostly
one characteristic time dominating the dynamics (i.e. the collision
time). In such systems, the only cause of (always very small)
deviations from usual FD relations is the non-Gaussian behavior of the
velocity statistics, but even in particular cases where velocity pdf
is strongly non-Gaussian Green-Kubo formulas may possibly work. The
fact that the granular temperature of the measured tracer is the
effective temperature is a quite obvious result if the original
derivation of Green-Kubo relations is followed, as there the effective
temperature appears to be simply $\langle v^2 \rangle$.

\section{Conclusions} 
\label{section:conclusion}

In this paper we have briefly reviewed the different approaches which
have been followed in the last few years in order to define a notion
of temperature for granular media. This question is a non-trivial one
for such systems where the usual notion of termodynamic temperature,
related to thermal agitation, does not play any obvious role. In these
systems the very possibility of consistently constructing a
thermodynamics is doubtful due to the fact that energy is lost through
internal friction, and gained by non-thermal sources such as tapping
or shearing.  The dynamical equations, whenever one could be able to
write them down explicitily, do not leave the microcanonical or any
other known ensemble invariant. Moreover these systems could never be
considered at equilibrium and even the existence of stationary states
is not always guaranteed. For instance often for dense granular media,
just as in the case of aging glasses, a stationary state cannot
be reached on experimental time scales.

Despite all these difficulties in the last few years there have been
several contributions which, though not yet completely satisfactory,
are interesting because they have opened a new perspective which is
worth pursuing in the future.

One of the key concept has been the notion of dynamic (or effective)
temperature, as defined in the framework of the
Fluctuation-Dissipation Theorem. Following the remarkable work done on
glassy systems where the partial violation of the
Fluctuation-Dissipation Theorem has been put in relation with the
existence of a so-called dynamical temperature (describing the slow
structural rearrangements of the system), a lot of work has been done
along the same lines in the framework of granular media.

Another key contribution is due to S.F. Edwards who put forward a very
ambitious approach to define a granular ``ensemble'' by looking at the
so-called blocked (or jammed) configurations.

In this paper we have tried to sum up (in a very partial and maybe
subjective way) all these efforts, classifying them with respect to
the different regimes a granular medium can be found in: the glassy
regime, the liquid-like and the granular gas one.

For the glassy-like regime in particular, we have reviewed Edwards'
approach and described a possible path to check its validity for two
non-mean field models: the Kob-Andersen model and the Tetris model.
From this and other studies it turns out that the notion of Edwards
compactivity seems to be closely related to that of dynamical
temperature. A somewhat surprising but very interesting result,
especially because it opens the way to experimental checks of Edwards'
hypothesis. For the liquid-like regime we have reported about recent
experimental results where an unusual ``thermometer'', in the specific
case a torsion pendulum, has been used to test the soundness of the
temperature concept in a continuously shaken container of tiny
beads. Also in this case the temperature has been defined in the
framework of the Fluctuation-Dissipation Theorem and its value seems
to be consistent with values of the so-called granular temperature,
defined in terms of the velocity fluctuations. The relation of this
temperature with the one defined in the glassy regime is an open
problem. Finally for the granular gas regime, we have reported about
the validity of the Fluctuation-Dissipation Theorem.  It turns out
that, while inelastic collisions produce large deviations with respect
to usual thermodynamics and hydrodynamics, linear response theory and
Fluctuation-Dissipation relations are still valid provided that the
``equilibrium temperature'' is replaced by the granular temperature of
each component of the gas. There exist deviations with respect to the
usual Green-Kubo formula which are due to non-Gaussian behavior of
velocity distribution (in particular for cooling granular gases) and
the non-stationarity due to the thermal cooling of the gas. These
deviations are indeed very small for driven granular gases.

The picture emerging is still partial, even in each specific
regime. It would be important to reinforce in the next years the
experimental research in order to check the theoretical predictions
and try to bridge some links between the different regimes, even
though we expect that the level of universality, for these
non-equilibrium systems, is very low.  The two extreme regimes seem
the most lacking. In particular in the glassy regime it would be
important to have some experimental check of Edwards's hypothesis. On
the other hand for granular gases there are already many predictions
which only call for an experimental check.

\section*{Acknowledgements} The authors wish to thank E. Bertin, 
E. Caglioti, V. Colizza, L. Cugliandolo, O. Dauchot, D. Dean,
J. Kurchan, M. Sellitto, E. Trizac, for many interesting discussions
and collaborations over the last few years. V.L. acknowledges the
hospitality of the Laboratoire de Physique Th\'eorique de
l'Universit\'e de Paris sud, Orsay, where this work has been
completed. A.P. acknowledges the support of a Marie Curie fellowship
under the contract n. MEIF-CT-2003-500944.

\end{document}